\documentclass[12pt]{article}
\usepackage{amsfonts}

\def\be{\begin{equation}}
\def\ee{\end{equation}}
\def\bea{\begin{eqnarray}}
\def\eea{\end{eqnarray}}

\def\otra{b}
\def\totra{{\tilde b}}

\def\1{\'{\i}}

\def\k{{\kappa}}

\def\dd{{\rm d}}
\def\>#1{{\mathbf#1}}

\def\funciona{{\cal G}}
\def\funcionb{{\cal F}}

\def\y{y}
\def\ji{z}
\def\la{\lambda}
\def\m{\mu}

 \def\otra{b}

\begin{document}

\thispagestyle{empty}

\hfill \

\ 
\vspace{0.5cm}

\begin{center} {\Large{\bf {Universal integrals for superintegrable systems on
}} }

{\Large{\bf{N-dimensional spaces of  constant curvature}}}

\end{center}

\medskip

\begin{center} \'Angel Ballesteros and Francisco J.
Herranz 
\end{center}

\begin{center} {\it {Departamento de F\1sica,  Universidad de Burgos, 
09001 Burgos, Spain}}
\end{center}

  \medskip

\begin{abstract} 
\noindent
An infinite family of classical superintegrable  Hamiltonians
defined on the $N$-dimensional spherical, Euclidean and hyperbolic spaces are shown to
have a common set of $(2N-3)$ functionally independent constants of the motion. 
Among them, two different subsets of $N$ integrals in involution (including the
Hamiltonian) can always be  explicitly identified. As particular cases, we recover in a
straightforward way most of the superintegrability properties of the 
Smorodinsky--Winternitz  and generalized Kepler--Coulomb systems on spaces of constant
curvature and we   introduce as well new classes of (quasi-maximally) superintegrable potentials on
these spaces. Results here presented are a consequence of the
$sl(2,\mathbb R)$ Poisson coalgebra symmetry of all the Hamiltonians, together with an
appropriate use of the phase spaces associated to Poincar\'e and Beltrami coordinates. 
\end{abstract}

\bigskip\bigskip\bigskip\bigskip



\newpage

\setcounter{equation}{0}

\renewcommand{\theequation}{\arabic{equation}}


\noindent

\section{Introduction}

An $N$-dimensional ($N$D) completely integrable Hamiltonian $H^{(N)}$ is called {\it
maximally superintegrable} (MS) if there exists a set of
$(2N-2)$ globally defined  functionally independent constants of the motion that
Poisson-commute with $H^{(N)}$. Among these constants, at least, two different subsets of
$(N-1)$ constants in involution can be found; these are connected with the
fact that the system must possess separable solutions to the Hamilton--Jacobi equation
in, at least, two different coordinate systems. From the dynamical point of view, all
bounded motions of MS systems are closed and strictly periodic.
In a similar way, if
$H^{(N)}$ has
$(2N-3)$ integrals with the abovementioned properties the system will be called  {\it
quasi-maximally superintegrable} (QMS). Obviously, all MS systems are QMS ones.  

The search of explicit MS (or QMS) systems is usually afforded by fixing the number
$N$ of degrees of freedom together with some assumptions concerning both
the functional dependence of
$H^{(N)}$ and of the integrals of the motion. In the case of natural systems with $H^{(N)}=T+U$,
the configuration space of the system (a space with a given  curvature, for
instance) is chosen through a fixed expression for the kinetic energy
$T$, and MS potentials $U$ are investigated by imposing the existence of a suitable
number of (unknown) integrals of the motion, that are usually assumed to be
quadratic in the momenta and have also to be found simultaneously. 

However, it turns out
that although for a given space and dimension there exists a certain number of
superintegrable potentials, only very few of them admit arbitrary
$N$D generalizations. For example, in the $N$D Euclidean space the only two
known MS Hamiltonians with integrals quadratic in the momenta are the oscillator
potential with $N$ centrifugal terms (the so called Smorodinsky--Winternitz system
\cite{fris,evans2}) and the generalized Kepler--Coulomb potential with $(N-1)$
centrifugal terms~\cite{Evansa}. 

The aim of this work is to face this problem from a quite different
viewpoint, which is based on introducing a common $sl(2,\mathbb
R)\otimes 
\dots^{N)}\otimes sl(2,\mathbb R)$ symmetry framework for a large family of
$N$D QMS systems with integrals quadratic in
the momenta. Surprisingly enough, all the systems introduced in this way will share,  by
construction, the same set of
$(2N-3)$ functionally independent integrals of the motion, which will be
explicitly given. Among them, only one constant of the motion $C^{(N)}$ will depend
on the $N$ degrees of freedom and the remaining ones will be functions of a
decreasing number of positions and momenta.

Therefore,
in this framework the Hamiltonian $H^{(N)}$ will be the only function that characterizes each
individual QMS system. In the case of natural systems, the kinetic
term
$T$ will give us the information concerning the configuration space, and the potential  term $U$
will fully identify the interactions. We will show that, by following this approach, QMS systems on
$N$D  spaces  of constant curvature can be easily understood and constructively
studied. In this respect, the geometric interpretation of the canonical coordinates
and momenta on each space turns out to be essential, and will be induced by the
Hamiltonian transcription of the Poincar\'e and Beltrami coordinates.

Moreover, we stress that for some specific choices of $H^{(N)}$ the systems here presented  are
indeed MS ones, and we will recover as particular cases the $N$D
Smorodinsky--Winternitz and generalized Kepler--Coulomb systems, that can now be
 understood as distinguished cases of an infinite family of systems with the same
underlying symmetry. For both cases the additional independent integral of the motion,
which is not provided by the underlying symmetry, can be found by other methods and will
be explicitly given.

The paper is organized as follows. In the next Section we present the main result that
defines the infinite family of QMS systems together with the explicit form of their
`universal' $(2N-3)$ integrals of the motion. A sketch of the proof of this theorem
is given,  which is based on the symmetry of our systems under an $sl(2,\mathbb
R)\otimes 
\dots^{N)}\otimes sl(2,\mathbb R)$ algebra (or, alternatively, in terms of an
$sl(2,\mathbb R)$ Poisson coalgebra~\cite{CP,BR}). Afterwards, many
different examples of  Hamiltonian systems belonging to this family (and thus sharing
the same set of universal integrals) are presented.  In Section 3 we construct QMS
systems on the $N$D Euclidean space
$\mathbb E^N$ by using a kinetic energy $T$ given in terms of the usual momenta conjugated to Cartesian coordinates,
that can be naturally identified with  
the canonical coordinates $\>q$. Some of these systems were already given
in~\cite{BR,Deform,CRMAngel} but others are here identified within the coalgebra
symmetry framework for the first time.  Section 4 is devoted to QMS systems living on
the
$N$D spherical
$\mathbb S^N$ and hyperbolic $\mathbb H^N$ spaces, for which  the coordinates
$\>q$ are   identified    either with
Poincar\'e or with Beltrami   coordinates coming, respectively, from  
stereographic  and  central projections on
$\mathbb R^{N+1}$.  Finally, some remarks and open problems are mentioned.


\section{An infinite family of QMS systems}

The main result of this paper can be stated as follows.

\medskip

 \noindent
{\bf Theorem 1.}
{\it Let $\{\>q,\>p \}=\{(q_1,\dots,q_N),(p_1,\dots,p_N)\}$ be $N$ pairs of
canonical variables. The $N$D Hamiltonian
\begin{equation}
H^{(N)}={\cal H}(\>q^2,\tilde{\>p}^2,\>q\cdot\>p)
\label{hgen}
\end{equation}
with ${\cal H}$ any smooth function and
\begin{equation}
 \>q^2=\sum_{i=1}^N q_i^2  \qquad\    \tilde{\>p}^2=
    \sum_{i=1}^N \left(  p_i^2+\frac{\otra_i}{ q_i^2} \right)
\equiv \>p^2 +  \sum_{i=1}^N  \frac{\otra_i}{ q_i^2}  \qquad\ \>q\cdot\>p
=
  \sum_{i=1}^N  q_i\, p_i
\label{qp}
\end{equation}
where $b_i$ are arbitrary real parameters, is 
quasi-maximally superintegrable. The   $(2N-3)$ functionally
independent and `universal' integrals of the motion are
\bea
&& C^{(m)}= \sum_{1\leq i<j}^m \left\{ ({q_i}{p_j} -
{q_j}{p_i})^2 + \left(
\otra_i\frac{q_j^2}{q_i^2}+\otra_j\frac{q_i^2}{q_j^2}\right)\right\}
+\sum_{i=1}^m \otra_i \nonumber\\
&& C_{(m)}= \sum_{N-m+1\leq i<j}^N \left\{ ({q_i}{p_j} -
{q_j}{p_i})^2 + \left(
\otra_i\frac{q_j^2}{q_i^2}+\otra_j\frac{q_i^2}{q_j^2}\right)\right\}
+\sum_{i=N-m+1}^N \otra_i 
\label{cinfm}
\eea
where $m=2,\dots, N$. Moreover, the sets of $N$ functions
$\{H^{(N)},C^{(m)}\}$ and $\{H^{(N)},C_{(m)}\}$ $(m=2,\dots, N)$ are in
involution. 
}
\medskip

Some remarks are in order:
\begin{itemize}

\item Note that $C^{(N)}=C_{(N)}$, so that the number of different integrals is  $(2N-3)$.

\item The underlying symmetry can be interpreted as a generalization of the
`spherical' one, since  
$ H^{(N)}={\cal H}(\>q^2,{\>p}^2,\>q\cdot\>p)
$ is recovered when all $b_i=0$. In this case,  the   constants of
the motion   (\ref{cinfm}) are just sums of squares of certain
angular momentum components
$ { l}_{ij}=q_ip_j-q_jp_i$. In particular, ${\cal C}^{(m)}=\sum_{1\leq i<j}^m {
l}^2_{ij}$ and
${\cal C}_{(m)}=\sum_{N-m+1\leq i<j}^N { l}^2_{ij}$.

\item As we shall see in the sequel, the additional centrifugal terms ${\otra_i}/{ q_i^2}$ in $H^{(N)}$
come from the freedom to add such a term in the corresponding symplectic
realization of an $sl(2,\mathbb R)$ Poisson algebra. These terms will be essential in order to
understand several known results concerning QMS systems on spaces of constant
curvature.
  
\end{itemize}

 \noindent
{\bf Proof.}
This relies on the fact that, for any choice of the
function
${\cal H}$, the Hamiltonian
$H^{(N)}$ has an $sl(2,\mathbb R)$ Poisson coalgebra symmetry~\cite{BR} (under a certain
symplectic realization) and, as a consequence, the integrals of the motion come from the left
and right
$m$-coproducts of the Casimir function for $sl(2,\mathbb R)$~\cite{Deform, CRMAngel}. Let us
summarize the main steps of this construction.

We recall that the
$sl(2,\mathbb R)$ Poisson coalgebra is given by the following Lie--Poisson brackets,
coproduct, Casimir and (primitive) coproduct  $\Delta$:
\begin{equation}
 \{J_3,J_+\}=2 J_+     \qquad  
\{J_3,J_-\}=-2 J_- \qquad   
\{J_-,J_+\}=4 J_3     
\label{ba}
\end{equation}
\begin{equation}
\begin{array}{l}
\Delta(J_l)=  J_l \otimes 1+ 1\otimes J_l \qquad l=+,-,3
\end{array}
\label{bb}
\end{equation}
\begin{equation} 
{\cal C}=  J_- J_+ -J_3^2  . 
\label{bc}
\end{equation} 
A  one-particle symplectic realization of (\ref{ba}) is given by  
\begin{equation}
 J_-=q_1^2     \qquad \qquad  
J_+=
      p_1^2+ {\otra_1}/{ q_1^2} \qquad   \qquad 
J_3=
    q_1\,p_1     
\label{symp}
\end{equation}
where $\otra_1$ is a real parameter that labels the representation through ${\cal
C}=\otra_1$. 
The coalgebra approach~\cite{BR} provides the corresponding
$N$-particle symplectic realization   through the $N$-sites coproduct of (\ref{bb})
living on
$sl(2,\mathbb R)\otimes  \dots^{N)}\otimes sl(2,\mathbb
R)$~\cite{Deform}:
\begin{equation}
 J_-=\sum_{i=1}^N q_i^2\equiv \>q^2 \quad\    J_+=
    \sum_{i=1}^N \left(  p_i^2+\frac{\otra_i}{ q_i^2} \right)
\equiv \>p^2 +  \sum_{i=1}^N  \frac{\otra_i}{ q_i^2}  \quad\ J_3=
  \sum_{i=1}^N  q_i p_i\equiv \>q\cdot\>p  
\label{be}
\end{equation}
where $\otra_i$ are $N$ arbitrary real parameters. This means that the $N$-particle generators
(\ref{be}) fulfil the  commutation rules (\ref{ba}) with respect to the   canonical Poisson
bracket 
$$
\{f,g\}=\sum_{i=1}^N\left(\frac{\partial f}{\partial q_i}
\frac{\partial g}{\partial p_i}
-\frac{\partial g}{\partial q_i} 
\frac{\partial f}{\partial p_i}\right).  
\label{bg}
$$

As a consequence of the coalgebra approach, these generators Poisson commute with the
$(2N-3)$ functions (\ref{cinfm}) given by the sets $C^{(m)}$ and
$C_{(m)}$, which are obtained from the `left' and `right'
$m$-th coproducts of the Casimir (\ref{bc}) with $m=2,3,\dots,N$  (see \cite{CRMAngel} for
details). If we label the $N$ sites on
$sl(2,\mathbb R)\otimes sl(2,\mathbb R)\otimes\dots\otimes sl(2,\mathbb R)$ by
$1\otimes 2\otimes\dots\otimes N$, the `left' Casimir  
 ${\cal C}^{(m)}$  is defined on the sites $1\otimes 2\otimes \dots \otimes m$, while
the `right' one ${\cal C}_{(m)}$  is defined on $m\otimes \dots \otimes N-1\otimes N$.
Moreover, it is straightforward to prove that the   $2N-2$ functions $\{ {\cal
C}^{(2)},{\cal C}^{(3)},\dots ,{\cal C}^{(N)}\equiv {\cal C}_{(N)}, {\cal
C}_{(N-1)},\dots {\cal C}_{(2)},{\cal H}
\}$ are functionally independent (assuming that ${\cal H}$ is not a function of ${\cal C}$) and the coalgebra symmetry ensures that each of the subsets
$\{{\cal C}^{(2)},\dots,{\cal C}^{(N)},{\cal H} \}$ or $\{{\cal C}_{(2)},\dots,{\cal
C}_{(N)},{\cal H} \}$ are formed by $N$  functions in involution~\cite{BR,CRMAngel}.

Therefore, any   arbitrary function ${\cal H}$ defined on the
$N$-particle symplectic realization of $sl(2,\mathbb R)$ (\ref{be}) is of the form  
(\ref{hgen}), that is, 
$$
{H}^{(N)}= {\cal H}\left(J_-,J_+,J_3\right)
={\cal H}\left(\>q^2,\>p^2+\sum_{i=1}^N\frac{\otra_i}{q_i^2},\>q\cdot\>p  \right)
$$
 and defines a QMS
Hamiltonian system. 

Notice that when $N=2$, the generic  function ${\cal H}$ determines  an {\em
integrable}   Hamiltonian as there is a single constant of the motion ${\cal
C}^{(2)}\equiv {\cal C}_{(2)}$. On  the contrary, when $N=3$ the Hamiltonian $\cal
H$ (\ref{hgen}) can be called minimally or `weak' 
{\em superintegrable} (by following the terminology of \cite{Evansa} for this specific
dimension) since it is endowed with three integrals $\{ {\cal C}^{(2)},{\cal
C}^{(3)},{\cal C}_{(2)}\}$. However, we remark that for arbitrary  
$N$ there is a single constant of the motion left  to assure   maximal
superintegrability. In this respect, we   stress that some specific choices of  
$\cal H$  comprise maximally superintegrable systems as well, but the remaining
integral does not come from the coalgebra symmetry and has to be
deduced by making use of alternative procedures.


\section{QMS systems on the Euclidean space}

It is immediate to realize that the $N$D  kinetic energy  on $\mathbb E^N$
directly arises through the generator
$J_+$ (\ref{be}) as it includes the usual momenta $\>p^2$. Then if we add  some
smooth   function
${\cal F}(J_-)$ as the potential term we can construct different QMS systems. Some
possibilities are the following.
\medskip

\noindent
$\bullet$ {\em The $N$D Evans system}:
\be
{\cal  H} 
=   \frac1{2m} J_+ +{\cal F}(J_-)
= \frac1{2m} 
\,  \>p^2+{\cal
F}(\>q^2)+\sum_{i=1}^N \frac{\totra_i}{2 q_i^2} 
\label{cb}
\ee
where $\totra_i=\otra_i/m$. This describes a particle of mass $m$   under the action
of a  central potential ${\cal F}(\>q^2)$  with
$N$  centrifugal terms associated with the
$\totra_i$'s. The $N=3$ case was introduced in~\cite{Evansa}.

\medskip

\noindent
$\bullet$ {\em The Smorodinsky--Winternitz  
 system}~\cite{fris,evans2,Deform,evans3,grosche1,Kalninsa}.  By choosing ${\cal
F}(J_-)=\omega^2 J_-$ in the above Hamiltonian, we recover a system   
formed by an isotropic harmonic oscillator of mass $m$ with 
 angular frecuency
$\omega$ together with $N$ centrifugal barriers:
\be
{\cal  H} 
=   \frac1{2m} J_+ +\omega^2 J_-
= \frac1{2m} 
\,  \>p^2+\omega^2 \>q^2+\sum_{i=1}^N \frac{\totra_i}{ 2  q_i^2} .
\label{cc}
\ee
 This Hamiltonian is known to be maximally superintegrable. A set of $N$ additional
constants of the motion (which do not come from the coalgebra symmetry) is
\be
{\cal I}_i=p_i^2+2 m \omega^2  q_i^2+m  {\totra_i}/{q_i^2}\qquad i=1,\dots,N.
\label{ccc}
\ee
Each constant ${\cal I}_i$   is
functionally independent with respect to both the set  (\ref{cinfm}) and ${\cal  H}$, thus it
can be taken as the `lost' integral.
\medskip

\noindent
$\bullet$ {\em A  
Garnier-type system}~\cite{ Garnier,Wojciechowski}. 
If we set ${\cal F}(J_- )=\omega^2 J_- +
\delta J_-^2$, with real parameter $\delta$,   we find a
degenerate Garnier  Hamiltonian (with 
 $a_i=\omega^2$), which besides the previous harmonic  term also comprises
 quartic oscillators: 
\be
{\cal  H} 
=   \frac1{2m} J_+ +\omega^2 J_-+\delta J_-^2
=\frac1{2m} 
\,  \>p^2+\omega^2 \>q^2+\delta\>q^4+\sum_{i=1}^N \frac{\totra_i}{ 2 q_i^2}   .
 \label{cd}
\ee
The generalization to even-order nonlinear oscillators is straightforward, since the generic
system
\be
{\cal  H} 
=   \frac1{2m} J_+ +\omega^2 J_-+\sum_{k=1}^\infty\delta_k J_-^{k+1}
=\frac1{2m} 
\,  \>p^2+\omega^2 \>q^2+\sum_{k=1}^\infty\delta_k \>q^{2(k+1)}+\sum_{i=1}^N \frac{\totra_i}{
2 q_i^2}   
 \label{cdinfty}
\ee
is a QMS one for any choice of the $\delta_k$ parameters.
\medskip

\noindent
$\bullet$ {\em A  
generalized Kepler--Coulomb system}~\cite{Evansa}. 
The choice  ${\cal F}(J_- )=- k J_-^{-1/2}$, with real
constant
$k$, gives rise to the superposition of the   Kepler--Coulomb potential with $N$
centrifugal barriers:
\be
{\cal  H} 
=   \frac1{2m} J_+ -k  J_-^{-1/2}
=\frac1{2m} 
\,  \>p^2-\frac{k}{ \sqrt{\>q^2 }} +\sum_{i=1}^N \frac{\totra_i}{ 2 q_i^2}   .
 \label{ce}
\ee
Such a Hamiltonian is known to be maximally superintegrable under the condition that,
at least, one of the centrifugal terms vanishes. In particular, if  a single
$\totra_i=0$, an additional constant of the motion (and functionally
independent with respect to  (\ref{cinfm}) and (\ref{ce})) arises. This additional integral  is found to be 
\be
{\cal L}_i=\sum_{l=1 }^N p_l(q_l p_i-q_i p_l)+ k m\,\frac{q_i}{\sqrt{\>q^2}}- m
\sum_{l=1;l\ne i}^N \totra_l\,\frac{ q_i}{ q_l^2}  .
\label{cee}
\ee
Likewise, if another $\totra_j=0$ ($j\ne i$),  the function ${\cal L}_j$ can be proven to be also a
constant of the motion, and so on. When all the   $\totra_i$'s vanish the system
reduces to the  proper Kepler--Coulomb potential  and the $N$ additional constants of
the motion
${\cal L}_i$ are the components of the  Laplace--Runge--Lenz $N$-vector on
$\mathbb E^N$.

\medskip
\noindent
$\bullet$ {\em Stationary electromagnetic fields}. Certain momenta-dependent potentials
can also be  obtained within this framework through an additional term depending on the
generator
$J_3$.  Let us consider
\bea
&&{\cal  H} 
=   \frac1{2m}\, J_+ -\frac e m\, J_3\,\funciona(J_-)+ e
\funcionb(J_-)\nonumber \\[2pt]
&&\qquad  = \frac 1{2m}\, \>p^2-\frac e m\,
(\>p\cdot\>q) \funciona(\>q^2)+ e \funcionb(\>q^2)  +\sum_{i=1}^N \frac{\totra_i}{ 2
q_i^2}  
 \label{cf}
\eea
where $e$ is a real parameter and $\funciona(J_-)$ is a smooth function. When $N=3$,
the above Hamiltonian describes  a particle with   mass $m$ and charge $e$ that moves
 on
$\mathbb E^3$   under the action of an electromagnetic field, that is, ${\cal
H}=\frac 1{2m}(\>p-e\>A)^2+e\psi$,  where the (time-independent)   vector   $\>A$ and
  scalar  $\psi$ potentials are given by
$$ 
\psi(\>q)=\funcionb(\>q^2)   -\frac{e}{2m}\,\>q^2 \funciona^2(\>q^2) 
 +\sum_{i=1}^3 \frac{\totra_i}{ 2 e q_i^2}  
\qquad \>A(\>q)= \>q \,\funciona(\>q^2).
\label{cg}
$$
The electric  $\>E=-\nabla \psi$ and magnetic $ \>H=\nabla\times \>A$  fields
turn out to be
$$ 
  \>E=\left(\frac e m \, \funciona^2+ \frac{2e}{m}\, \>q^2 \funciona
\funciona^\prime-2\funcionb^\prime    \right)\>q  +\frac 1 e \biggl(\frac{\totra_1}{ 
q_1^3},\frac{\totra_2}{ 
q_2^3},\frac{\totra_3}{ 
q_3^3} \biggr)
\qquad \>H=0 
\label{ch}
$$
where $\funciona^\prime$ and $\funcionb^\prime$ are
the derivatives with respect to their variable $\>q^2$.   
Recall that 2D integrable  electromagnetic Hamiltonians have
been studied in \cite{Hietarinta,Dorizzi,winter}.
 We  stress that this kind of construction  
can also be  applied  to obtain 
$N$D QMS Fokker--Planck Hamiltonians  (see
\cite{Hietarinta} and references therein).
\medskip

\noindent
$\bullet$ {\em Systems with coordinate-dependent mass}. If we multiply the
former kinetic term $J_+$ in the above Hamiltonians by an arbitrary smooth function
${\cal M}(J_-)$, we obtain systems with a variable mass   depending on $\>q^2$ such
as, for instance,
\be
{\cal  H} 
=   \frac1{2 {\cal M}(J_-)} \,J_+ +{\cal F}(J_-)
= \frac1{2{\cal M}(\>q^2)} 
\,  \>p^2+{\cal
F}(\>q^2)+\sum_{i=1}^N \frac{\otra_i}{2{\cal M}(\>q^2) q_i^2} .
\label{ci}
\ee

We stress that 
the expression (\ref{ci}) can be  alternatively interpreted as a QMS
system on `some' space  which would be determined through the metric coming from the
  kinetic energy  $J_+/{\cal M}(J_-)$. In particular, in the next Section we will show
how certain specific Hamiltonians of the type (\ref{ci}) will define superintegrable
systems on the sphere $\mathbb S^N$   and the hyperbolic space $\mathbb H^N$.


\section{QMS systems on the sphere and the hyperbolic space}
 
Recall that for a constant sectional curvature $\k$, both the $N$D
sphere $\mathbb S^N$
$(\k>0)$ and the $N$D hyperbolic space
$\mathbb H^N$
$(\k<0)$ can be embedded in an ambient linear space $\mathbb R^{N+1}$ with
Weiertrass coordinates $(x_0,\>x)=(x_0,x_1,\dots,x_N)$ fulfilling the `sphere'
constraint $\Sigma$: $x_0^2+\k \>x^2=1$. The metric on the proper $N$D spaces is given
by~\cite{VulpiLett,CRMVulpi}:
\be
\dd
s^2= {1\over\k}
\left(\dd x_0^2+\k  \dd \>x^2\right)\biggr|_{\Sigma} 
\label{ccii}
\ee
where $ \dd \>x^2= \sum_{i=1}^N \dd x_i^2$. The $N+1$ ambient coordinates can be
parametrized in terms of
$N$ intrinsic coordinates in different ways. 

Firstly, we consider the stereographic
projection~\cite{Doub}   from the ambient coordinates
$(x_0,\>x)\in
\Sigma$ to the Poincar\'e ones
$\>\y\in \mathbb R^N$ with pole  
$(-1,\>0)\in \mathbb R^{N+1}$. Hence $(-1,\>0)+\la\,
(1,\>\y)\in\Sigma$ and we obtain  that
\be
\la=\frac{2}{1+  \k\>\y^2}
\qquad x_0=\la -1=\frac {1-\k\>\y^2}{1+
\k\>\y^2}\qquad 
\>x=\la\>\y=\frac{2\>\y}{1+
\k\>\y^2}.
\label{cj}
\ee
Secondly, we   apply the central projection   from the Weiertrass coordinates  to the
Beltrami ones
$\>\ji\in \mathbb R^N$ with pole  
$(0,\>0)\in \mathbb R^{N+1}$. Thus $(0,\>0)+\m\,
(1,\>\ji)\in\Sigma$ and we find  
\be
\m=\frac{1}{\sqrt{1+ \k \>\ji^2}}\qquad
 x_0=\m\qquad 
\>x=\m\, \>\ji=\frac{\>\ji}{\sqrt{1+ \k \>\ji^2}}.
\label{ck}
\ee
Then the metric (\ref{ccii}) in Poincar\'e and Beltrami coordinates  turns out to be
\be
\dd s^2=4\,\frac{\dd \>\y^2}{(1+\k\>\y^2)^2}=\frac{(1+\k\>\ji^2)\dd\>\ji^2-\k
(\>\ji\cdot \dd\>\ji)^2}{(1+\k\>\ji^2)^2} .
\label{cl}
\ee
 Next if we write the metrics as Poincar\'e
${\cal T}^{\rm P}$ and Beltrami ${\cal T}^{\rm B}$ free Lagrangians
\be
{\cal T}^{\rm P}= \frac{m\dot\>y^2}{2(1+\k\>y^2)^2}\qquad
{\cal T}^{\rm B}=\frac{m}{2}\left( \frac{(1+\k\>\ji^2)\dot\>\ji^2-\k
(\>\ji\cdot \dot\>\ji)^2}{(1+\k\>\ji^2)^2}
\right)
\label{da}
\ee
the   momenta $\>p_y$ and $\>p_z$ conjugated to $\>y$ and $\>z$ can be deduced
\be
\>p_y= \frac{m\dot\>y}{(1+\k\>y^2)^2}\qquad \>p_z= m \left(
\frac{(1+\k\>\ji^2)\dot\>\ji-\k (\>\ji\cdot \dot\>\ji)\>\ji}{(1+\k\>\ji^2)^2}\right)
 .
\label{db}
\ee

Therefore, the previous expressions show that the $N$D kinetic energy in both Poincar\'e
${\cal T}^{\rm P}$ and Beltrami ${\cal T}^{\rm B}$ phase spaces can be written as the
following functions of the symplectic realization (\ref{be}) of the generators of
$sl(2,\mathbb R)$:
\begin{equation}
\begin{array}{l}
\displaystyle{ {\cal T}^{\rm P}=\frac{1}{2m}\left( 1+\k J_-\right)^2 J_+=
\frac{1}{2m}\left( 1+\k \>q^2\right)^2 \>p^2}\\[8pt]
\displaystyle{ {\cal T}^{\rm B}=\frac{1}{2m}\left( 1+\k J_-\right)\left(  J_+ +\k
J_3^2\right)=
\frac{1}{2m}(1+\k \>q^2)\left( \>p^2+\k (\>q\cdot \>p)^2 \right) }
\end{array}
\label{dd}
\end{equation}
provided that all $\otra_i=0$, $(\>q,\>p)\equiv (\>y,\>p_y)$ for ${\cal
T}^{\rm P}$ and $(\>q,\>p)\equiv (\>z,\>p_z)$ for ${\cal
T}^{\rm B}$. When the $\otra_i$'s are taken as arbitrary parameters,
the $N$ additional curved `centrifugal'  potentials arising in (\ref{dd}) are
written in ambient coordinates in a very simple form:
\be
 \sum_{i=1}^N \frac{\otra_i}{2m \y_i^2}\,(1+\k
\>y^2)^2=2\,\sum_{i=1}^N\frac{\totra_i}{x_i^2}\qquad   \sum_{i=1}^N
\frac{\otra_i}{2m \ji_i^2}(1+\k
\>\ji^2)=\sum_{i=1}^N\frac{\totra_i}{2 x_i^2}.
\label{df}
\ee
It is worth remarking that (\ref{df}) is actually a direct consequence of the
symplectic realization of $sl(2,\mathbb
R)$ (\ref{be}) with $\otra_i\ne 0$. This, in turn,  explains why the famous
`centrifugal' terms
do  appear everywhere in the classifications of superintegrable systems on the three
classical Riemannian spaces~\cite{Evansa,RS,PogosClass1,PogosClass2}.

Recall as well that the radial (geodesic polar)
distance $r$ from an arbitrary point to the origin in $\mathbb S^N$ and   
$\mathbb H^N$ along the geodesic joining both points is
expressed in terms of Weiertrass~\cite{VulpiLett,CRMVulpi},  Poincar\'e and Beltrami
coordinates as
\be
\frac{1}{\k}\tan^2(\sqrt{\k}\,r)
=\frac{\>x^2}{x_0^2}=\frac{4\>\y^2}{(1-\k\>\y^2)^2}=\>\ji^2.
\label{de}
\ee
Note that in the constant curvature analogues of the oscillator and
Kepler--Coulomb problems the Euclidean radial distance is just
replaced by the function
$ 
\frac{1}{\sqrt{\k}}\tan(\sqrt{\k}\,r).
$ 

Therefore, by taking into account (\ref{dd})--(\ref{de}) we finally present some QMS and MS
Poincar\'e
${\cal H}^{\rm P}$ and Beltrami
${\cal H}^{\rm B}$ Hamiltonians  that are constructed by adding some suitable functions depending on $J_-$ to
(\ref{dd}) and by considering arbitrary centrifugal terms $\otra_i$'s. These systems are the curved counterpart of
the Euclidean systems (\ref{cb})--(\ref{cee}) which simultaneously cover 
$\mathbb S^N$  
($\k>0$),
$\mathbb H^N$    ($\k<0$), and $\mathbb E^N$    ($\k=0$; in this flat limit 
$r^2= {\>x^2} = {4\>\y^2} =\>\ji^2$). We stress that, again, all of them share
the {\em same} set of constants of the motion  (\ref{cinfm}), although in all these
curved cases the geometric meaning of the canonical coordinates and momenta is
completely different to the Euclidean one, as explained above. To make this geometrical interpretation more explicit, for each system we shall give both the Poincar\'e and Beltrami versions of each Hamiltonian.
\medskip

\noindent
$\bullet$ {\em A curved Evans system}. It would be given by
\begin{equation}
\begin{array}{l}
\displaystyle{ {\cal H}^{\rm P}={\cal T}^{\rm P}+{\cal F} \left(   \frac{4
J_-}{ (1-\k J_-)^2 } 
\right) =
\frac{\left( 1+\k \>q^2\right)^2 \>p^2}{2m}}+{\cal F}\left(
\frac{4\>q^2}{(1-\k\>q^2)^2} \right) + \sum_{i=1}^N\frac{2\totra_i}{x_i^2}
\\[8pt]
\displaystyle{ {\cal H}^{\rm B}={\cal T}^{\rm B}+ {\cal F}\left(J_-\right)=
\frac{1}{2m}(1+\k \>q^2)\left( \>p^2+\k (\>q\cdot \>p)^2 \right) +{\cal F}\left(
\>q^2\right)+\sum_{i=1}^N\frac{\totra_i}{2 x_i^2}}.
\end{array}
\label{eb}
\end{equation}

\noindent
$\bullet$ {\em The curved Smorodinsky--Winternitz  
 system}~\cite{VulpiLett,CRMVulpi}. 
Such a system is just the Higgs oscillator~\cite{Higgs,Leemon} (that arises as either
polynomial or rational potential in these coordinates) plus the corresponding centrifugal
terms:
\begin{equation}
\begin{array}{l}
\displaystyle{ {\cal H}^{\rm P}={\cal T}^{\rm P}+   \frac{4 \omega^2 
J_-}{ (1-\k J_-)^2 } =
\frac{\left( 1+\k \>q^2\right)^2 \>p^2}{2m}}+ 
\frac{4\omega^2 \>q^2}{(1-\k\>q^2)^2}   + \sum_{i=1}^N\frac{2\totra_i}{x_i^2}
\\[8pt]
\displaystyle{ {\cal H}^{\rm B}={\cal T}^{\rm B}+\omega^2 J_- =
\frac{1}{2m}(1+\k \>q^2)\left( \>p^2+\k (\>q\cdot \>p)^2 \right) + \omega^2 
\>q^2+\sum_{i=1}^N\frac{\totra_i}{2 x_i^2}}.
\end{array}
\label{ec}
\end{equation}
This Hamiltonian is  maximally superintegrable and the remaining
constant of the motion can be taken from any of the following
$N$ functions (to be compared with (\ref{ccc})):
\begin{equation}
\begin{array}{l}
\displaystyle{ {\cal I}_i^{\rm P}=\left( p_i(1-\k \>q^2) + 2\k (\>q\cdot \>p) q_i 
\right)^2+ \frac{8 m \omega^2 q_i^2}{(1-\k\>q^2)^2}   +m  {\totra_i}\,
\frac{(1-\k\>q^2)^2} { q_i^2}  }\\[8pt]
\displaystyle{ {\cal I}_i^{\rm B}=\left( p_i+\k (\>q\cdot \>p) q_i \right)^2+2 m
\omega^2  q_i^2+m  {\totra_i}/{q_i^2}\qquad i=1,\dots,N. }\\[2pt]
\end{array}
\label{cccx}
\ee

\noindent
$\bullet$ {\em A  curved
Garnier-type system}. By following the same prescription, QMS quartic curved 
oscillators can be defined as
\begin{equation}
\begin{array}{l}
\displaystyle{ {\cal H}^{\rm P}={\cal T}^{\rm P}+   \frac{4 \omega^2 
J_-}{ (1-\k J_-)^2 }+ \frac{16 \delta 
J_-^2}{ (1-\k J_-)^4 } }\\[8pt]
\displaystyle{ \qquad\qquad =
\frac{\left( 1+\k \>q^2\right)^2 \>p^2}{2m}}+ 
\frac{4\omega^2 \>q^2}{(1-\k\>q^2)^2} + 
\frac{16\delta \>q^4}{(1-\k\>q^2)^4}   + \sum_{i=1}^N\frac{2\totra_i}{x_i^2}
\\[8pt]
\displaystyle{ {\cal H}^{\rm B}={\cal T}^{\rm B}+\omega^2 J_- +\delta J_-^2 =
\frac{(1+\k \>q^2)\left( \>p^2+\k (\>q\cdot \>p)^2 \right) } {2m}+ \omega^2 
\>q^2+\delta \>q^4+\!\sum_{i=1}^N\frac{\totra_i}{2 x_i^2}}.
\end{array}
\label{ed}
\end{equation}
The expressions for the curved QMS analogues of the higher-order nonlinear oscillators (\ref{cdinfty}) are straightforward.

\noindent
$\bullet$ {\em A  curved
generalized Kepler--Coulomb system}~\cite{RS,PogosClass1,PogosClass2,Schrodingerdual,
Schrodingerdualb,kiev}. 
The    curved Kepler--Coulomb potential with $N$ centrifugal terms 
corresponds to
\begin{equation}
\begin{array}{l}
\displaystyle{ {\cal H}^{\rm P}={\cal T}^{\rm P}-k \left(\frac{4  
J_-}{ (1-\k J_-)^2 } \right)^{-1/2}=
\frac{\left( 1+\k \>q^2\right)^2 \>p^2}{2m}}- k 
\frac{(1-\k\>q^2)}   {2  \sqrt{\>q^2}}+ \sum_{i=1}^N\frac{2\totra_i}{x_i^2}
\\[8pt]
\displaystyle{ {\cal H}^{\rm B}={\cal T}^{\rm B} -k  J_-^{-1/2}=
\frac{1}{2m}(1+\k \>q^2)\left( \>p^2+\k (\>q\cdot \>p)^2 \right) -\frac{k}{
\sqrt{\>q^2 }}+\sum_{i=1}^N\frac{\totra_i}{2 x_i^2}} .
\end{array}
\label{ee}
\end{equation}
This is again  a MS system provided that, at least, one   $\totra_i=0$. In
this case the remaining   constant of the motion  reads (compare with
the Euclidean version (\ref{cee})):
\bea
&&\!\!\!\!\!\!\!\!\!\!\!\!\displaystyle{ {\cal L}_i^{\rm P}=\sum_{l=1 }^N \left(
p_l(1-\k
\>q^2) + 2\k (\>q\cdot
\>p) q_l 
\right) (q_l p_i-q_i p_l) 
+ \frac{k m q_i}{2\sqrt{\>q^2}}- m
\sum_{l=1;l\ne i}^N \totra_l\,\frac{ q_i(1-\k \>q^2)}{ q_l^2}  }\nonumber\\
&&\!\!\!\!\!\!\!\!\!\!\!\!
\displaystyle{ {\cal L}_i^{\rm B}=\sum_{l=1 }^N  \left(
p_l+\k (\>q\cdot \>p) q_l \right)   (q_l p_i-q_i p_l)  
 +\frac{ k mq_i}{\sqrt{\>q^2}}- m
\sum_{l=1;l\ne i}^N \totra_l\,\frac{ q_i}{ q_l^2}  }.
\label{ceex}
\eea
If another $\totra_j=0$, then ${\cal L}_j^{\rm P,B}$ is also a
new constant of the motion. In this way the proper curved
Kepler--Coulomb system~\cite{Schrodinger} (with   all the  
$\totra_i$'s equal to zero) is obtained, and in that case (\ref{ceex}) are 
the
$N$ components of the  Laplace--Runge--Lenz vector on
$\mathbb S^N$ $(\k>0)$ and $\mathbb H^N$ $(\k<0)$.

More details on the latter Hamiltonians  and
their generalization to Lorentzian metrics will be given elsewhere. 
On the other hand, we remark that the definition of QMS systems on
spaces of  variable curvature can be achieved either by considering more general kinetic
energy terms or by making use of coalgebra deformations, that have been already shown to
underly the superintegrability on some variable curvature (Riemannian and relativistic) spaces~\cite{plb,jpa2D}.
Finally, the study of several interesting classes of non-natural Hamiltonians included in
(\ref{hgen}) are also worthy to be addressed in the future.  


\section*{Acknowledgements}

{This work was partially supported  by the Ministerio de Educaci\'on y
Ciencia   (Spain, Project FIS2004-07913) and  by the Junta de Castilla y
Le\'on   (Spain, Project VA013C05).}


\end{document}